\documentclass{article}
\usepackage{graphicx}

\newcommand{\deltat}{$\Delta t$}
\begin{document}

\title{Design and Performance of a Custom ASIC Digitizer for Wire Chamber Readout in 65\,nm CMOS Technology}

\author{
MyeongJae~Lee$^a$\thanks{Corresponding author.}, 
David~N.~Brown$^a$,
Jessica~K.~Chang$^a$ $^b$, \\
Dawei Ding$^a$ $^b$,
Dario~Gnani$^a$,
Carl~R.~Grace$^a$, \\
John~A.~Jones$^a$, 
Yury~G.~Kolomensky$^a$ $^b$,
Henrik~von~der~Lippe$^a$, \\
Patrick~J.~Mcvittie$^a$,
Matthew~W.~Stettler$^a$~ and \\
Jean-Pierre~Walder$^a$ \\
\llap{$^a$}Lawrence Berkeley National Laboratory, \\
Berkeley, California 94720, USA \\
\llap{$^b$}University of California at Berkeley, \\
Berkeley, California 94720, USA \\
}
\maketitle

\begin{abstract}
We present the design and performance of a prototype ASIC digitizer for
integrated wire chamber readout, 
implemented  in 65\,nm commercial CMOS technology. 
Each channel of the 4-channel prototype is composed of
two
16-bit Time-to-Digital Converters (TDCs), one 8-bit Analog-to-Digital Converter (ADC), a front-end preamplifier and shaper, plus digital and
analog buffers that support a variety of digitization chains.
The prototype has a multiplexed digital backend that executes a state machine,
distributes control and timing signals, and buffers data for serial output.
Laboratory bench tests measure the absolute TDC resolution between 74\,ps and 480\,ps,
growing with the absolute delay,
and a relative time resolution of 19\,ps.
Resolution outliers due to cross-talk between clock signals and supply or reference voltages are seen.
After calibration, the ADC displays good linearity and noise
performance, with an effective number of bits of 6.9. 
Under normal operating conditions the circuit consumes 32\,mW per channel.
Potential design improvements to address the resolution drift and tails are discussed.
\end{abstract}


\section{Introduction}
Wire chamber detectors require readout electronics
to amplify, shape, and digitize input signals from each wire, 
and digital circuitry to buffer and transmit the data.  Typically, these functions
are performed using a combination of separate ICs and discrete components.
Performing all these functions in a single IC would simplify and consolidate
the readout process.  Use of high-density CMOS processes can greatly reduce the
per-channel size and power requirements of the readout electronics, which can be
valuable for the high channel count detectors common in High Energy Physics (HEP) experiments. 

Dedicated ASICs for wire chamber signal amplification and shaping \cite{ref:Carioca, ref:NINO}, pulse-height
digitization \cite{ref:ILC_ADC, ref:SAR_ADC}, and time digitization \cite{ref:elefant, ref:Mester_TDC} have been previously
developed.  The ASIC developed for the BaBar experiment \cite{ref:elefant} combined time and pulse-height digitization.
These chips each required more than 50\,mW/channel, resulting in well over 100\,mW/channel
for a full electronics chain.  No single ASIC performs all the functions required for wire chamber signal processing
and digitization.

In this report, we present the design and performance of a prototype ASIC digitizer that provides
all the functionality required for wire chamber readout.  Each channel of the 4-channel prototype
contains a preamplifier, a discriminator, two TDCs, and one ADC, providing a full readout circuit
for both ends of a single wire.
In addition, each channel provides input and output buffers for digital and pre-amplified analog signals,
which can be multiplexed into the digitization chain, allowing a variety of signal path
configurations.  The ASIC has a minimal digital backend to buffer and transmit the
recorded data.

This ASIC was developed as a prototype of the digitizer 
for the straw tracker of the Mu2e experiment~\cite{ref:CDR}.
Mu2e is a proposed experiment to search for Charged Lepton Flavor Violation in muon to electron conversion in the field of a nucleus. 
The Mu2e tracker must measure the momentum of roughly $100\,{\rm MeV}/c$
electrons with a precision and accuracy better than $250\,{\rm keV}/c$ to achieve its science goals. 
The tracker design includes approximately 20,000 straws, operating in
vacuum inside a 1\,T solenoid, and arranged in many overlapping
layers, filling a roughly 
cylindrical volume 3 meters long by 1.4 meters in diameter.
Because of space limitations and thermal requirements specific to
operating the detector in vacuum,
the signal processing electronics must be compact and consume very low power.

To meet these requirements, we chose
to implement our ASIC in
65\,nm commercial CMOS technology, which had not been previously used for HEP applications. 
Choosing this technology also reduced engineering costs, through sharing design resources with
similar projects being developed at the same time.

\begin{figure}[!htb]
\begin{center}
\includegraphics[width=0.9\textwidth]{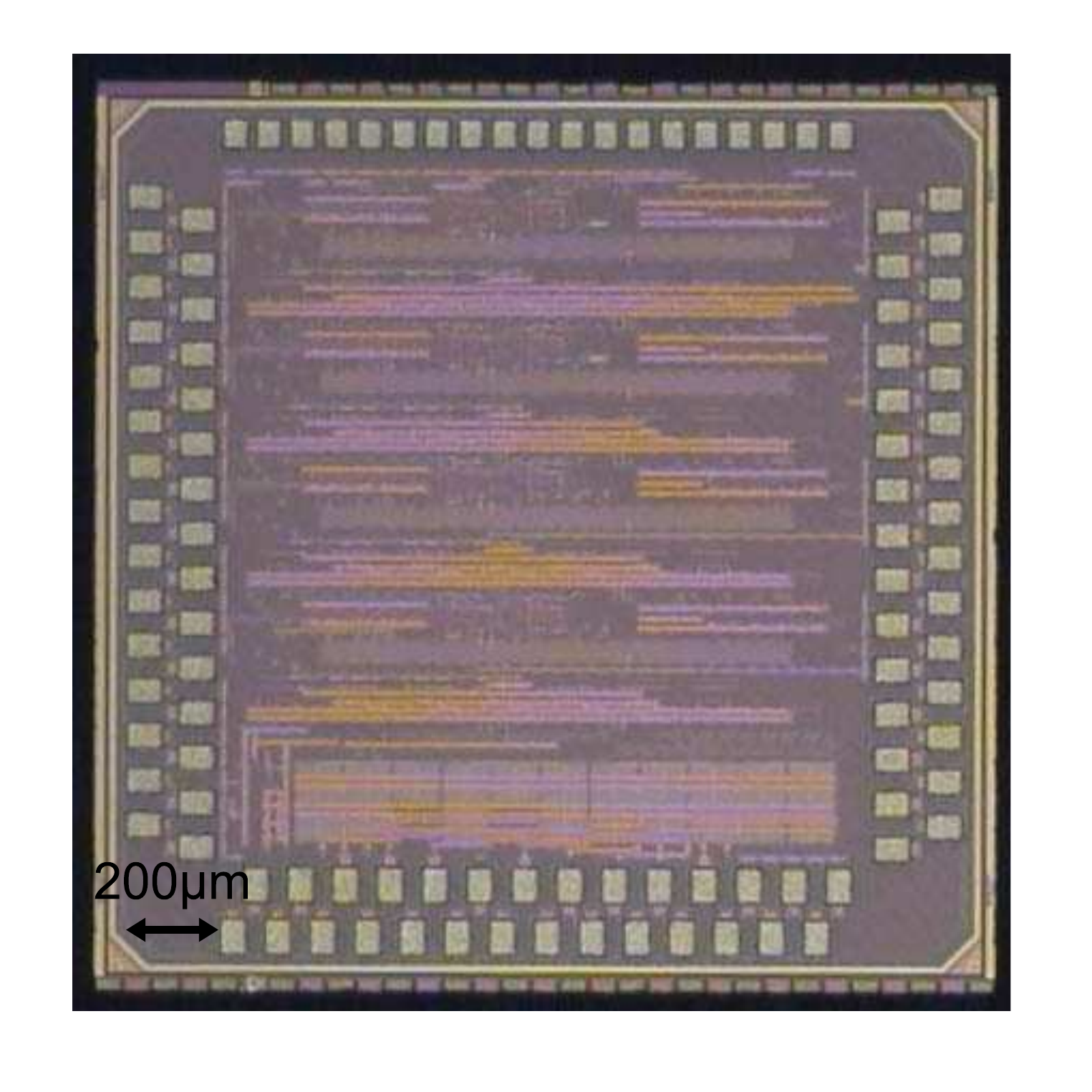}
\caption{Photo micrograph of the POM ASIC digitizer prototype.  The upper four repeating sections are the 4 POM channels, the bottom structure is an unrelated circuit.  POM is readout through the left, top, and right pad rows.}
\label{fig:pom_pic}
\end{center}
\end{figure}

We use the acronym POM (Processor Of Muon decays) to refer to our prototype ASIC, which is
shown in Fig. \ref{fig:pom_pic}.
The design concepts and functional blocks of POM are described in Sec. \ref{sec:design}.
The performance of the blocks are described 
in Sec. \ref{sec:performance}. 
The tolerance of POM to neutron radiation, and power consumption measurements, are also described in Sec. \ref{sec:performance}.

\section{Design of POM}
\label{sec:design}

The POM prototype was developed to test the concept of a self-contained wire chamber
digitizer, for the specific environment of the Mu2e experiment.
Detailed simulations of the Mu2e detector using \textsc{Geant4}
\cite{ref:Geant4} as well as
a complete implementation of the reconstruction and analysis chain were used to
set specific requirements for
POM, as summarized in Table \ref{table:spec}.
Many of the underlying requirements are not unique to Mu2e, so POM could be useful for other
applications in future.

Simulation studies show that efficient pattern recognition in Mu2e requires localization 
of charge deposition in 3 dimensions~\cite{ref:TDR}.  As with all wire detectors, 
the wire position, plus a measurement of the charge drift time, can determine the 
location of the charge deposition in the direction orthogonal to the wire. 
For the proposed Mu2e gas (Ar/CO$_{2}$ 80/20), adequate transverse position
resolution is obtained with $ < 500\,\rm{ps}$ drift time resolution.
The position along the wire can be determined by comparing the signal arrival time
at both ends of the wire.  Given typical values for the signal propagation time along the wire,
a time difference resolution of $\sim$50\,ps is required to achieve the
required $\sim$ 1\,cm longitudinal position resolution. 
Each POM channel provides drift time and time difference measurements
using dual TDCs operating with a common clock.
This allows POM to make 
precision time difference measurements without having to distribute 
fast clock signals over large distances.

The Mu2e environment produces many background particles, which can generate up to 1\,MHz of signals on a single wire.
To maintain high efficiency, the electronics deadtime (return to baseline) must be less than 200\,ns.  This implies a fast signal shaping time of roughly 20\,ns. 

Mu2e simulation studies also show that distinguishing signals generated by electrons from those generated by
protons or muons is
needed to reduce backgrounds to acceptable levels.  This separation can be achieved by measuring the
energy deposition per unit length, which depends on particle species.  POM provides a precise measurement
of the energy deposition using an ADC to digitize the wire current.
To separate pulses from nearly concurrent particles, 
multiple ADC waveform samples are used.
With a shaping time of ~20\,ns, the optimum 2-pulse separation can be made using 8 samples taken
at 50\,MHz sampling.

The key design challenges to implement mixed-signal circuits in 65\,nm are increased interconnect delay
and increased susceptibility to crosstalk.
To deal with the increased interconnect delay, we paid special attention to the floor-planning, especially in the TDC which required the tightest timing.  By minimizing the physical distance between communicating blocks and balancing the parasitics of matched lines, we were able to keep the timing skew under control.  Susceptibility to crosstalk 
can ideally be
minimized by bypassing supplies on-chip, regulating low-noise supplies internally, and generating key reference voltages internally.  The high cost of prototype production runs in 65nm resulted in severe die area constraints, 
which forced voltage supplies and regulator off-chip, increasing the susceptibility of POM to crosstalk.

\begin{table}[!htb]
\caption{The design goals for POM}
\begin{center}
\begin{tabular}{l c}
\hline
Parameters & Design goals  \\
\hline
Time-difference resolution & $< 50\,\rm{ps}$ \\ 
Absolute time resolution & $< 500\,\rm{ps}$ \\ 
Timing dynamic range & $> 1.7 \mu\rm{s}$ \\ 
ADC resolution & Minimum 7 bits (effective)\\ 
ADC sampling rate & $\ge 50 \rm{MS/s}$ \\ 
Input impedance & $50 \Omega$ (External preamp) / $300 \Omega$ (Internal preamp) \\ 
Baseline restoration time & 200\,ns (internal preamplifier) \\ 
Power dissipation & $< 50 \rm{mW/channel}$ \\ 
\hline
\end{tabular}
\end{center}
\label{table:spec}
\end{table}

\begin{figure}[!htb]
\begin{center}
\includegraphics[width=0.9\textwidth]{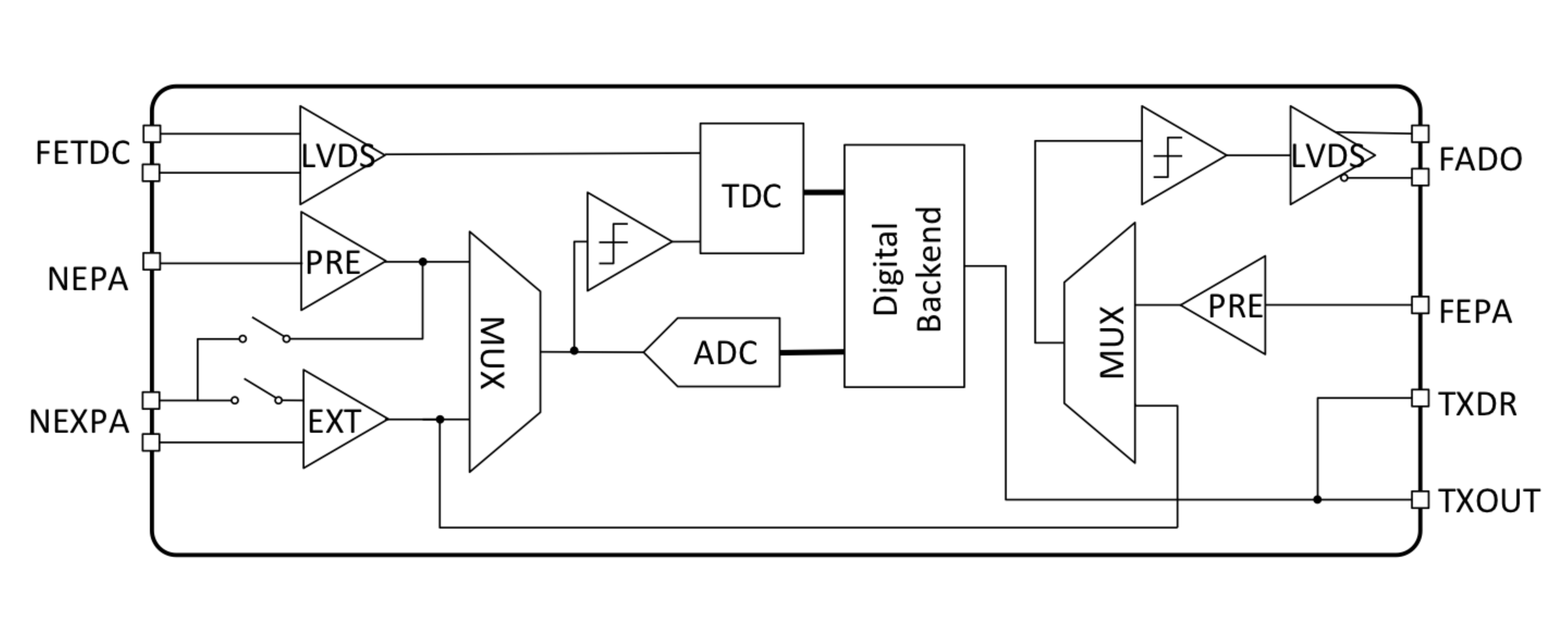}
\caption{Top level functional diagram of two POM channels.
}
\label{fig:pom_top}
\end{center}
\end{figure}

\begin{figure}[!htb]
\begin{center}
\begin{tabular}{c}
\includegraphics[width=0.9\textwidth]{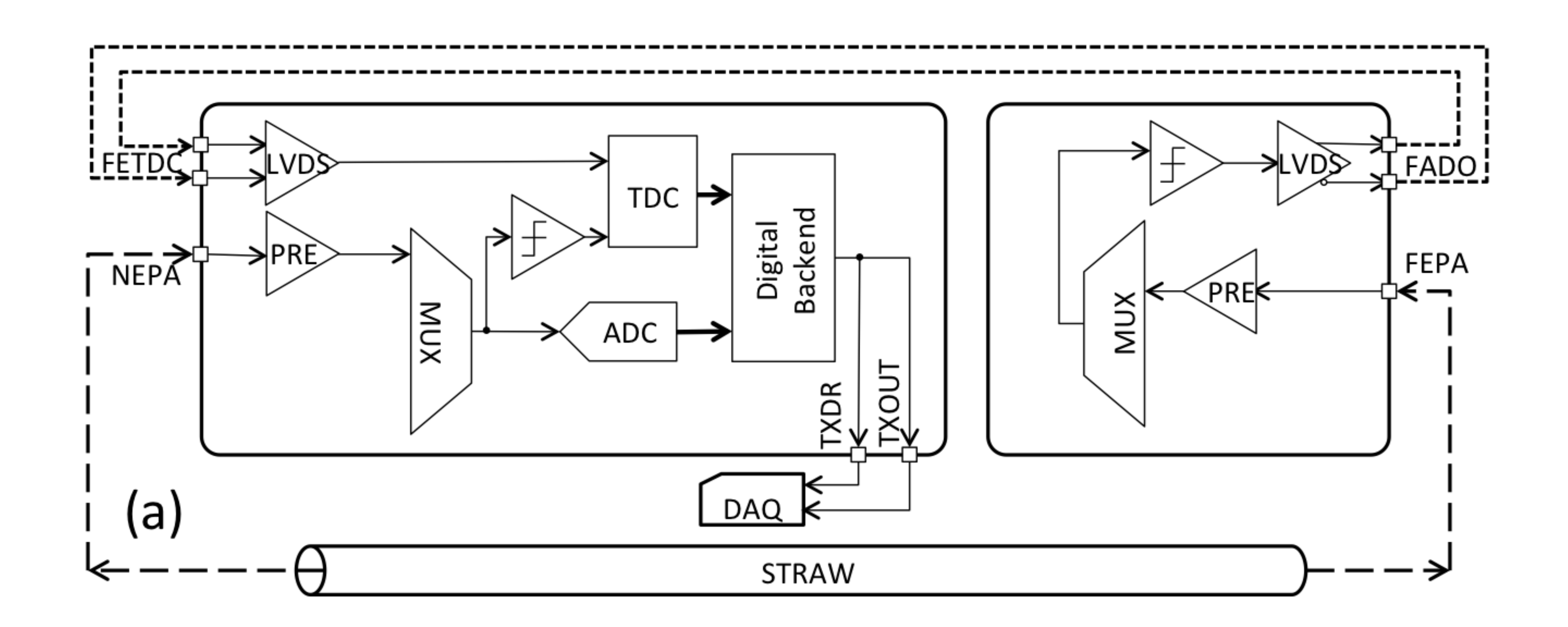} \\
\includegraphics[width=0.9\textwidth]{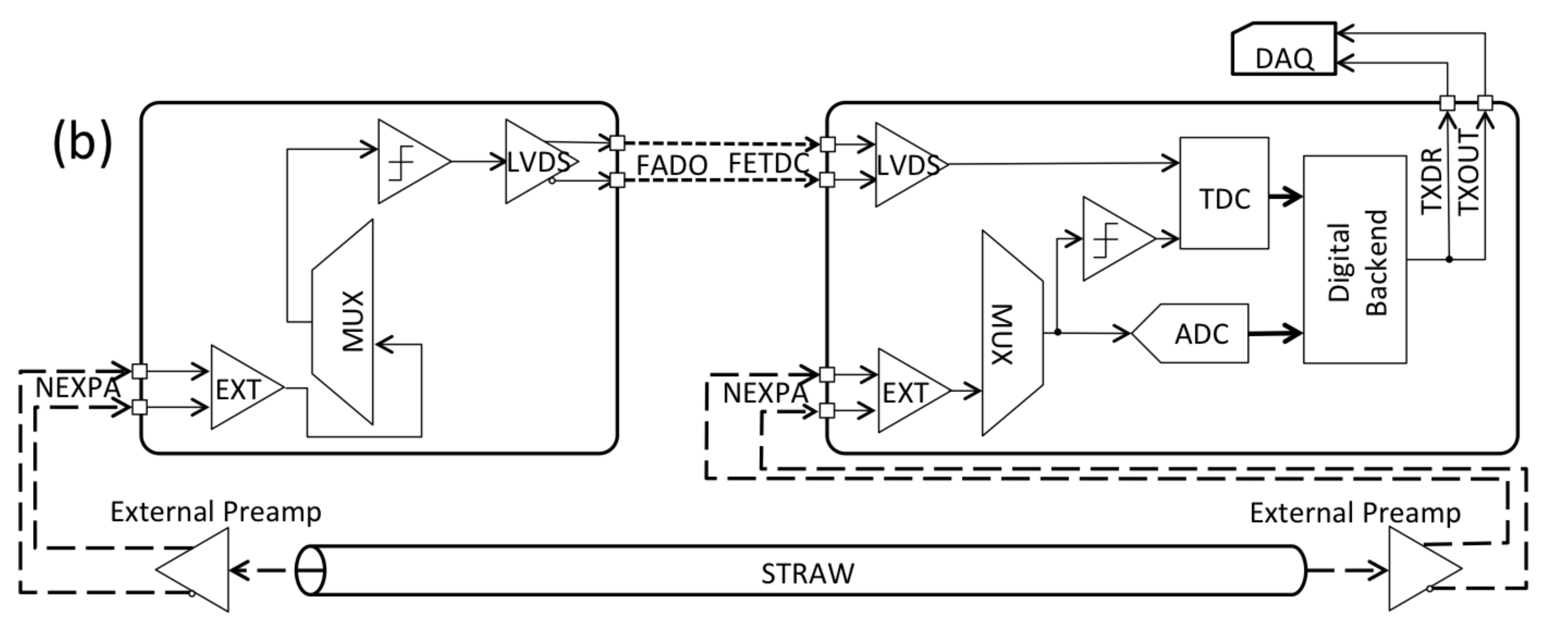} 
\end{tabular}
\caption{Two examples of POM configuration, (a) using internal preamplifier, and (b) using external preamplifier. The short and long dashed lines represent on-board connection between POM channels and analog signal line from a straw or external preamplifiers, respectivey. The arrows represent the direction of signal flow. The parts that are not used in each configuration are not shown for the purpose of clarity. } 
\label{fig:pom_connection}
\end{center}
\end{figure}

Figure \ref{fig:pom_top} shows the top level functional diagram of POM,
and Fig. \ref{fig:pom_connection} shows
a sample connection for a single straw.  The POM prototype can be
configured to accept
three different types of input signals: 
(1) a current pulse from an actual straw or emulator (depicted in Fig. \ref{fig:pom_connection}-(a)), 
(2) a voltage pulse from an external preamplifier or pulse generator (depicted in Fig. \ref{fig:pom_connection}-(b)), and 
(3) digital timing and analog voltage level test signals.  
The test signals can be fed through dedicated ports (not shown in Fig. \ref{fig:pom_top}) directly to
the TDC and ADC, allowing standalone tests of those components.
A challenge of time difference readout is bringing together signals from opposite straw ends,
which can be separated by large distances.  POM allows several configurations
that allow various location of POM around straw and the choice of external preamplifier,
which increase the flexibility of opposite-end time signal comparison, as shown in Fig. \ref{fig:pom_connection}. 

When configured for internal amplification (Fig. \ref{fig:pom_connection}-(a)), current signals from the near and far ends of the wire 
are connected to the NEPA (Near End Pre-Amplifier)  
and FEPA (Far End Pre-Amplifier) input ports, respectively. 
The current-sensitive preamplifier (PRE) amplifies and shapes the signal from NEPA,
and supplies an input voltage signal to the discriminator and the ADC.  The discriminator
output is used to stop the near end TDC.  
The far end current signal going to FEPA is also amplified, discriminated, and routed to FADO (Far-end Amplified Discriminated Output) ports as a Low-Voltage Differential Signal (LVDS). 
The output of FADO is routed to FETDC (Far End TDC input), and used to stop the far end TDC.  
Note that the circuits 
(blocks in Fig. \ref{fig:pom_connection})
used to digitize the wire ends can 
locate in a single POM channel, or 
be split between different POM channels, in either the same or different chips, allowing a variety of physical connection patterns.  By dividing the readout between
two POM chips, the analog circuitry can
be kept close to the wire ends, with only digital timing signals connecting the ends.  This minimizes the potential for 
noise pickup along the (long) signal path between the wire ends.

When configured for external amplification (Fig. \ref{fig:pom_connection}-(b)), voltage signals from external preamplifiers
attached to near and far wire ends are routed through NEXPA (Near End eXternal Pre-Amplifier) of two
different POM channels.
A receiver buffer (EXT) amplifies and shapes the signals from these ports.
As in the internal amplification configuration, separating the signal discrimination
across two POM channels allows locating the
POM chips close to the signal wire ends.

Each POM channel has an 8 bit pipeline ADC, designed to have a 1\,mV LSB (Least Significant Bit), and to operate at a maximum 65 MS/s sampling rate.
Each TDC digitizes the time difference 
between the threshold crossing of the pulse event and the (common) TDC reset signal. 
This reset signal is fed through another port which is not shown in the figure.
The design LSB of the TDC is 35\,ps, with 16 bits covering the maximum observable time in Mu2e of approximately 2 $\mu$s.

The discriminator (triangular block with a threshold symbol in Fig.~\ref{fig:pom_top}) generates a digital voltage pulse
while the shaped pulse exceeds its user-defined external threshold.  
The leading edges of the discriminator pulse signals provide the stop
signals for the TDCs.  

The digitized information goes through the Digital Backend. A digital state machine 
identifies events, packages them, responds to commands from the off-chip DAQ system, and serially transmits event data.
In the interest of simplicity and to reduce design costs, the digital backend of POM was designed to halt after
detecting a hit, and therefore the POM prototype provides no double-hit readout.
The designs of the preamplifier and discriminator, ADC, and TDC are described in detail in the following sections.

\subsection{Preamplifier and discriminator design}

\begin{figure}[!htb]
\begin{center}
\includegraphics[width=0.9\textwidth]{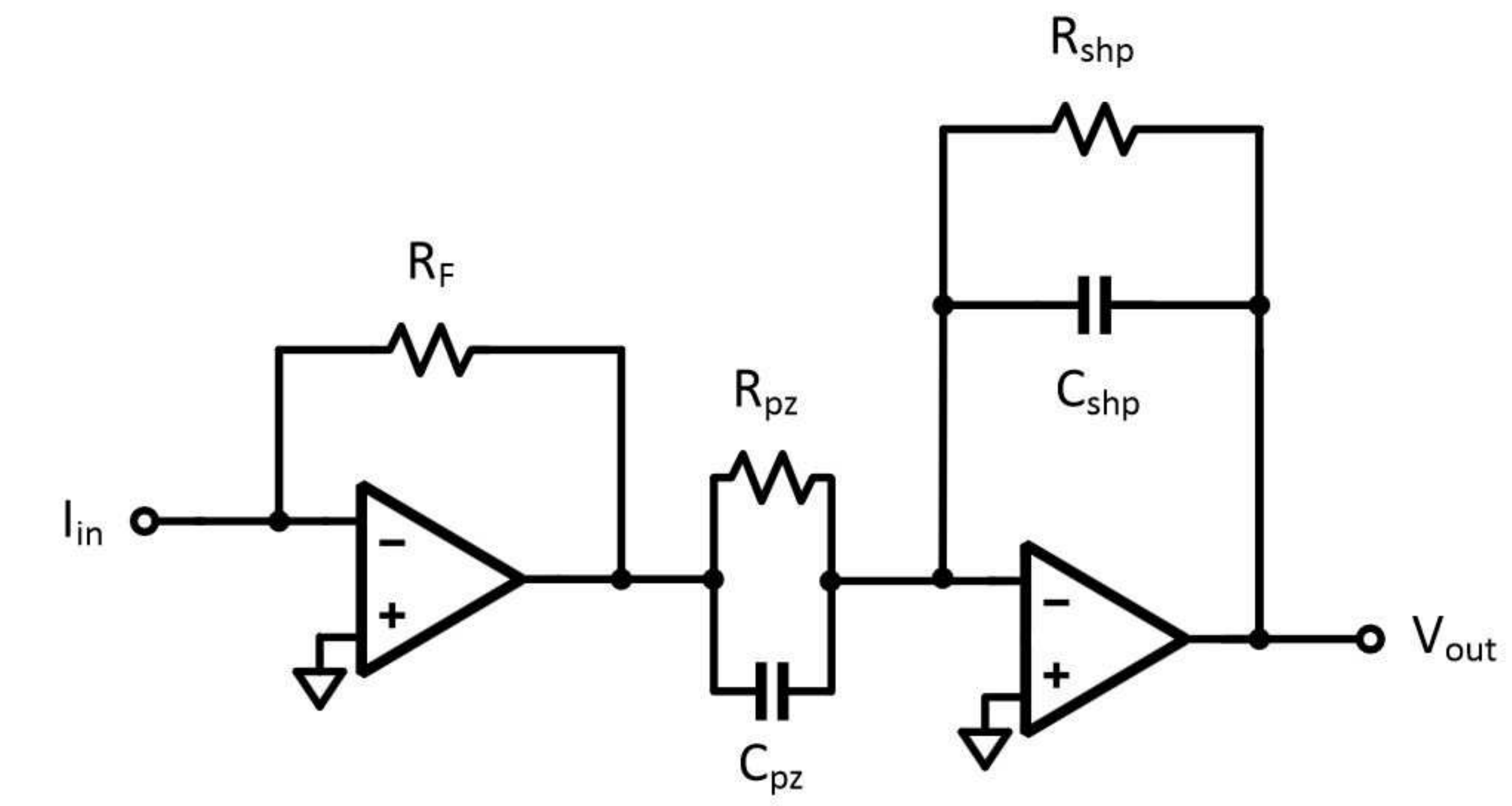}
\caption{POM frontend preamplifier design.}
\label{fig:frontend_design}
\end{center}
\end{figure}

The block diagram of the POM frontend preamplifier is depicted in Fig.~\ref{fig:frontend_design}. 
The circuit consists of a low-noise transimpedance amplifier followed by a first-order pole-zero cancellation network 
($\rm{R}_{\rm{PZ}}$ and $\rm{C}_{\rm{PZ}}$). 
A second transimpedance amplifier converts the current flowing through the pole-zero cancellation network into voltage and defines the signal fall time.  
Two configuration bits determine the gain, input impedance and dynamic range of the first transimpedance amplifier from 300\,mV/pC to 500\,mV/pC (at 100\,ns decay time) by changing the feedback resistor value ($\rm{R}_{\rm{F}}$),
while four configuration bits set the time constant of the pole-zero cancellation network by changing $\rm{R}_{\rm{PZ}}$. 
An analog multiplexer between the amplifier output and the discriminator allows configuring POM to connect either this amplifier or an external preamplifier to the discriminator (TDC) and the ADC.
An external reference voltage sets the discriminator threshold.
Simulation of the internal preamplifier found a
typical gain of $259 \sim 285\,{\rm mV/pC}$, Equivalent Noise Charge (ENC) of $1.5 \sim 1.8\,{\rm fC}$, 
and timing resolution of less than 30\,ps, for a 100\,fC input signal with decay constant of 100\,ns. 

\subsection{ADC design}
\label{sec:adc_design}

\begin{figure}[!htb]
\begin{center}
\includegraphics[width=0.9\textwidth]{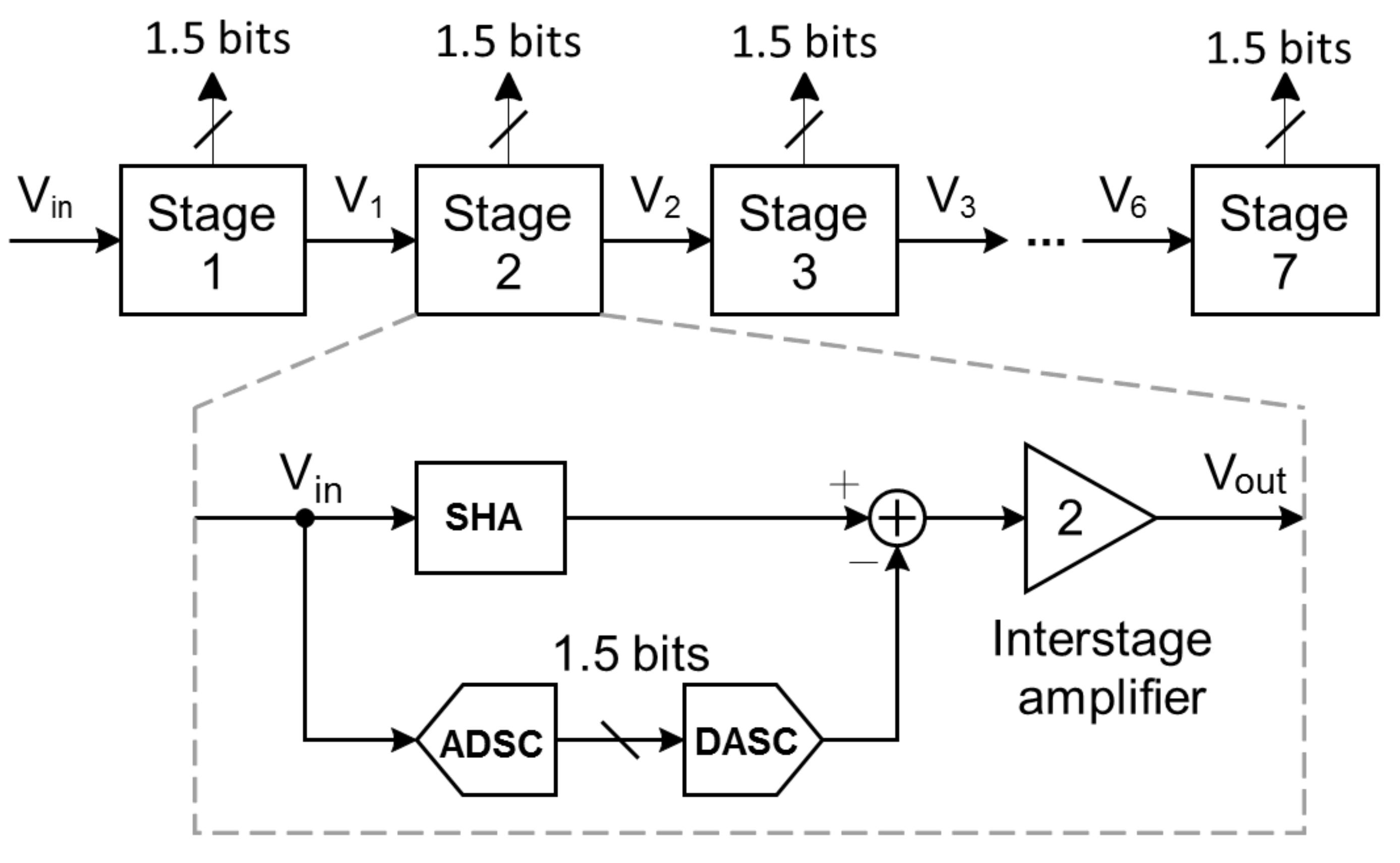}
\caption{POM ADC design.}
\label{fig:adc_design}
\end{center}
\end{figure}

The POM ADC is an 8 bit pipeline ADC, designed to operate at frequencies up to 65\,MHz.
The ADC design was adapted from an earlier ADC designed for Charged Coupled Device (CCD) readout~\cite{ref:HIPPO}.

Pipeline ADCs can provide both high speed and accuracy through the use of concurrent processing.  
The ADC accuracy depends on precision analog signal processing~\cite{ref:pipelineADC}.
In particular, high speed, high gain, and high accuracy operational amplifiers are required for interstage amplifiers. 
While Successive Approximation Register (SAR) ADCs can also achieve POM requirements in terms
of sampling rate and power dissipation \cite{ref:SAR_ADC}, they are not easily expandable to the number
of bits required for POM.

A block diagram of the ADC implemented in POM is shown Fig.~\ref{fig:adc_design}. 
It consists of 7 identical stages, each of which contains a low-resolution 1.5-bit 
\footnote{Two bits produced by each stage are combined in a combinatorial logic to form the ADC output, using effectively only 1.5 bits from each stage, to provide redundancy in the analog decision levels of the ADC. This redundancy can be used to make the converter tolerant to offset errors in its comparators and amplifiers. See Ref. \cite{ref:pipelineADC} for details.}
analog-to-digital subconverter (ADSC), 
a 1.5-bit digital-to-analog subconverter (DASC), an analog subtractor, a sample-and-hold amplifier (SHA), and a gain stage.  
In each stage, a coarse digital estimate of the analog input voltage is made by the ADSC, 
converted back to the analog domain by the DASC, and subtracted from the input voltage.  
The resulting residual voltage is
amplified by the interstage amplifier and passed to the next pipeline stage for finer conversion.    
The key feature of a pipeline ADC is that the SHAs allow each stage to operate concurrently on residues 
that correspond to different input samples, resulting in a high throughput 
that depends on the speed of a single stage and is almost independent of the number of stages.  

The POM ADC continuously digitizes its input and stores results to FIFO memory. 
When a pulse input triggers the discriminator, the FIFO data are captured so that two ADC samples preceding the
threshold crossing, 
and eight ADC samples following it are recorded. The ten ADC samples are collected into the data packet sent with the
rest of the digitization information.

Simulation studies incorporating the variation in capacitor and resistor
values specified by the foundry predicted a nonlinearity of the POM ADC
of less than 0.5 LSB.  Note however that the actual capacitor values used
in the POM ADC were substantially smaller than those used in the
foundry variation estimates.

\subsection{TDC design}

\begin{figure}[!htb]
\begin{center}
\includegraphics[width=0.9\textwidth]{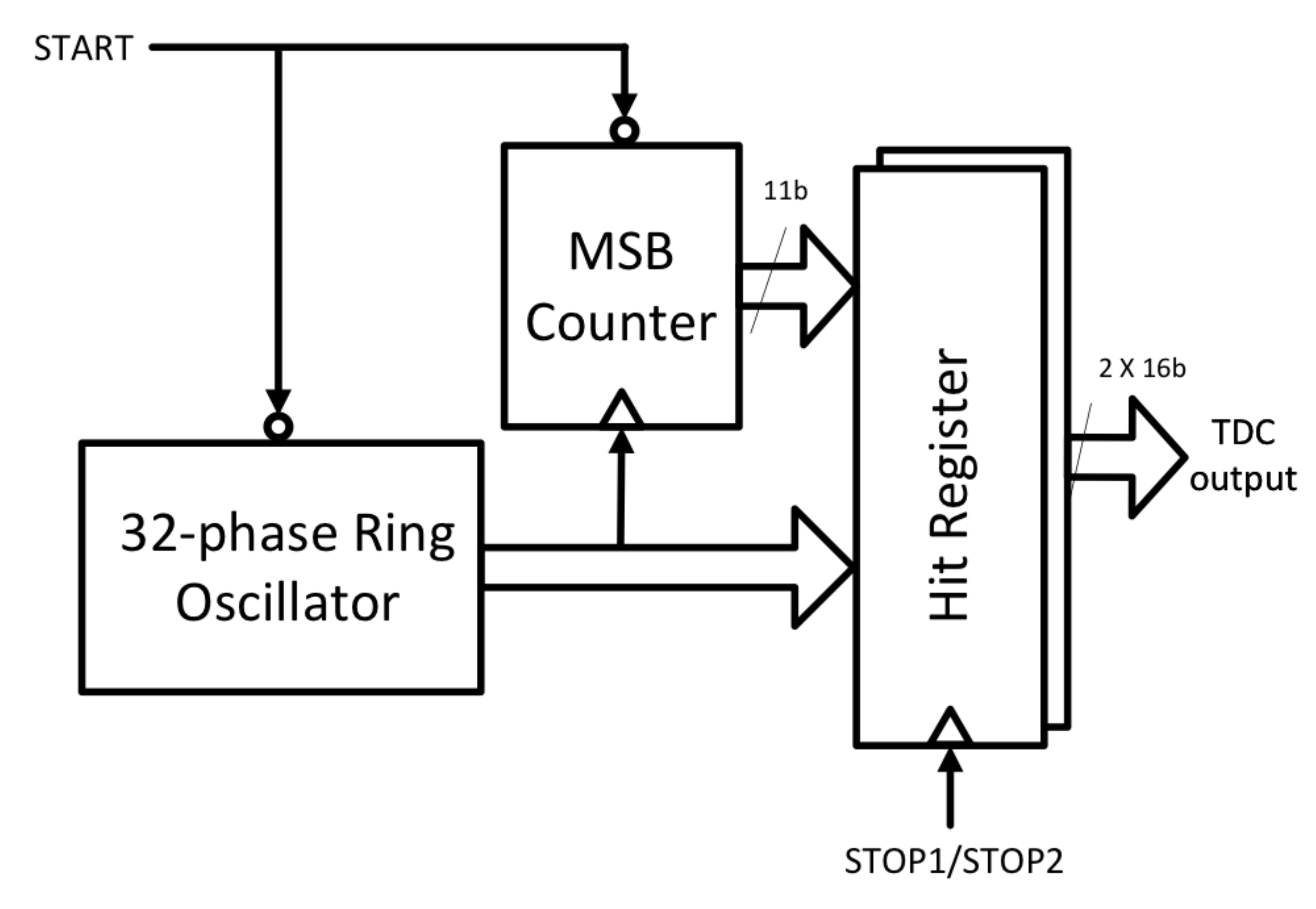}
\caption{Design of TDC of POM.}
\label{fig:tdc_design}
\end{center}
\end{figure}

A block diagram of the TDC is shown in Fig. \ref{fig:tdc_design}~\cite{ref:TDC}. 
A common 32-phase (16-stage) differential ring oscillator, running nominally at 850\,MHz
provides the 5 LSBs of the converter, and its core resolution.
A common 11-bit synchronous binary counter, counting the oscillator loops using one of its phases as a clock, 
extends the dynamic range to 16 bits, or 2 $\mu\rm{s}$.  The same ring oscillator is shared by both TDCs in the POM channel. 
This enables precision time difference measurements without relying
on clock distribution or synchronization between separate circuits. 
The ring oscillator speed is determined by the operation voltage and the temperature of the TDC, which were monitored during the entire POM testing period. Considering the measured dependence of TDC performance to the temperature and oscillator voltage, we could not find such large fluctuation that may affect to the speed of ring oscillator.

Two banks of high-speed latched sense amplifiers provide differential edge-sensitive capture and storage 
of the ring oscillator state (5 LSBs).
Another two banks of standard-cell D flip-flops capture and store the counter state (11 MSBs) at the stop signals' leading edges.
Re-synchronization circuits are designed to
process the stop signals and generate separate latching signal for the two types of register banks, 
which avoids metastability (and corresponding accuracy loss) in the counter registers.

Three fast digital signals are provided from the Phase-Locked Loop (PLL) of FPGA for the TDC time resolution measurement,
where one of them is used as a start signal of the TDC, and the other two are used as two TDC stop test inputs.
Using the special test input ports of the test board, the TDC can be directly started and stopped. 
This allows measuring the intrinsic timing performance, 
independent of the frontend analog signal processing. 

Detailed circuit simulation of the TDC predicted an LSB of 37\,ps, when operated using
the nominal ring oscillator voltage at a temperature of $ 30\,^{\circ}\mathrm{C}$.
Note that a PLL  to synchronize
the ring oscillator with the POM ADC clock,
in order to stabilize the TDC dependence to temperature,
was not included by design, 
because
a PLL was not necessary to meet the TDC design requirements.

\section{Performance Evaluation}
\label{sec:performance}
Two different custom test boards were designed to evaluate POM performance. 
The first test board mounted one POM chip, and connected to a commercial FPGA evaluation board 
through a 132 pin FPGA Mezzanine Card (FMC) connector to send control and data signals.
The evaluation board has a Virtex-5 FPGA, which controls and 
processes the data from POM. It communicates with a Linux-based DAQ
system via the UDP protocol. 
This test board is used for performance evaluation of a single POM chip.

The second test board was a prototype motherboard using POM for
wire chamber readout.  It contains  four POM chips, and uses an
on-board Kintex-7 FPGA for signal processing, control, 
and environment monitoring features. 
The second test board was designed to read out up to 8 straws, and for use in radiation tolerance tests. 
The results from both test boards are described in this section.  

The test board receives two types of inputs and distributes them to POM: test mode inputs and analog mode inputs.
The test mode inputs are used to test the ADC and TDC 
without the effects from the analog electronics. 
The analog inputs can be from a real straw detector, or a similar signal generated by an external pulse generator. 

The operation clock of POM test was 50\,MHz throughout this paper, except the ADC performance evaluation which was carried in 25\,MHz, to resolve the non-linearity issue (described below).

\subsection{ADC Performance}

\begin{figure}[!htb]
\begin{center}
\includegraphics[width=0.9\textwidth]{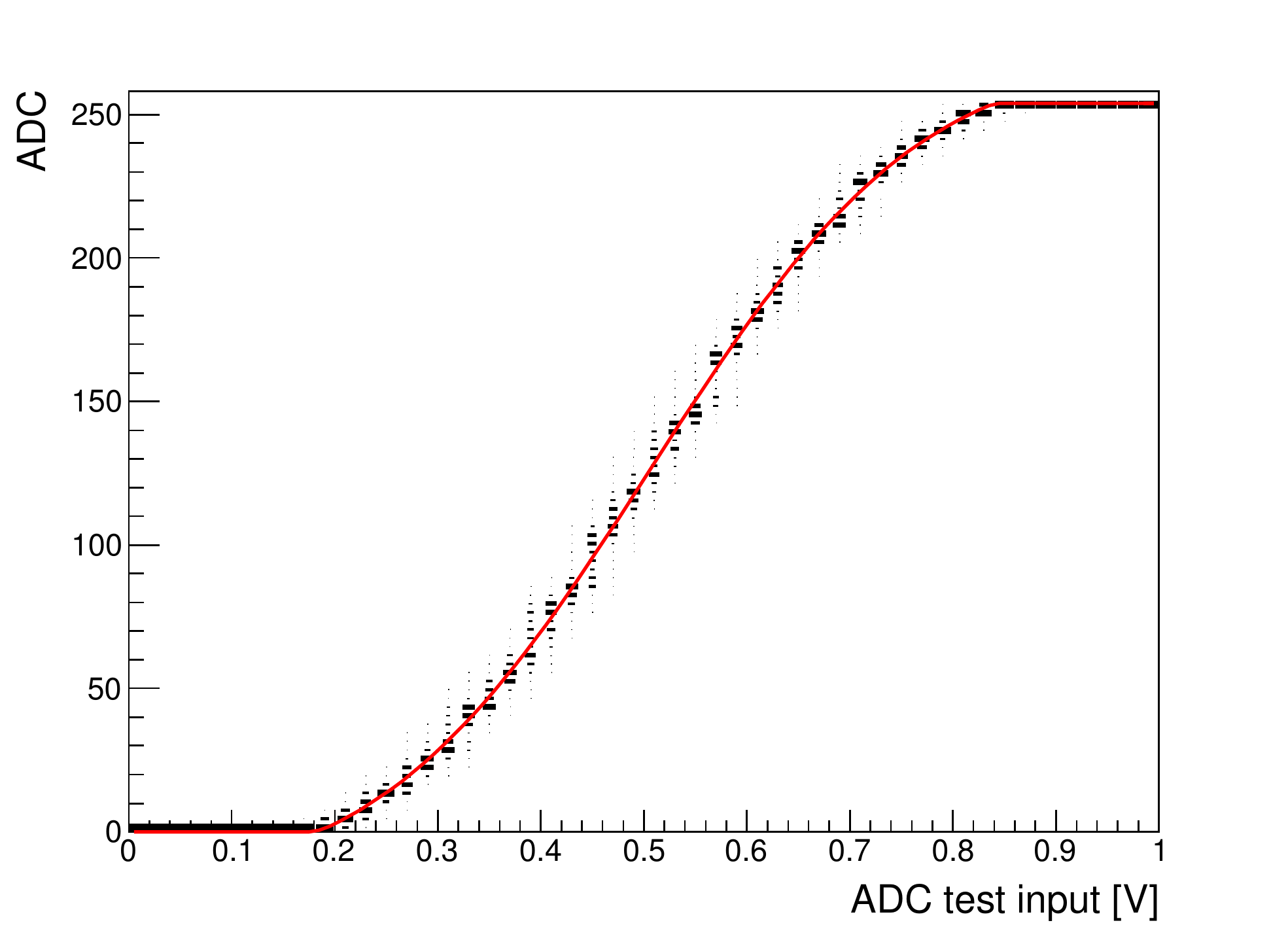}
\caption{ADC response to DC test inputs. The black boxes and the red line represent the ADC data and their fit to a truncated error function, respectively.}
\label{fig:adc_lin}
\end{center}
\end{figure}

\begin{figure}[!htb]
\begin{center}
\includegraphics[width=0.9\textwidth]{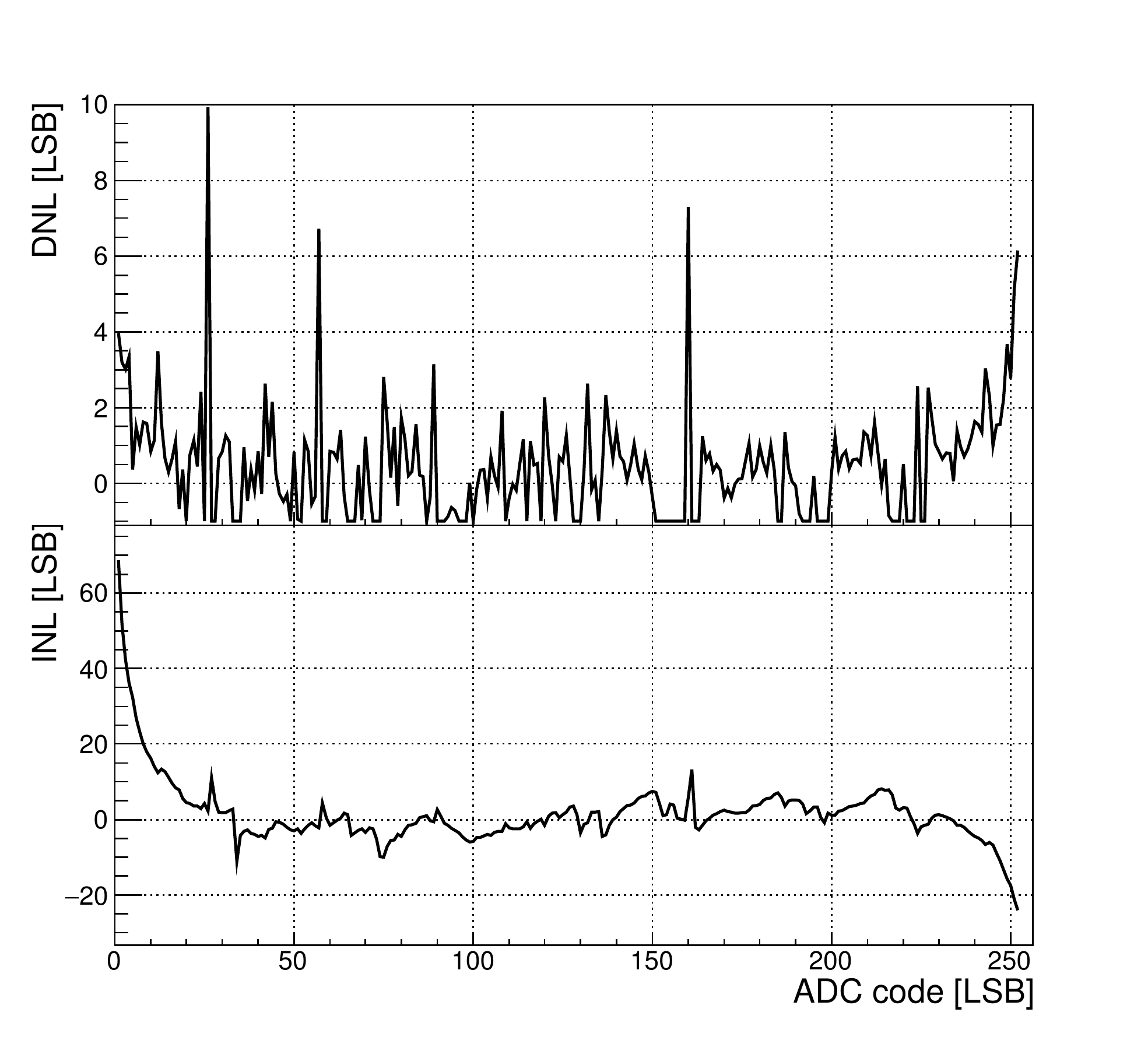}
\caption{The Differential Non-linearity (DNL, top plot) and Integral Non-linearity (INL, bottom plot) of POM ADC.}
\label{fig:adc_dnlinl}
\end{center}
\end{figure}

The linearity and the noise level of the ADC was evaluated by scanning the test input voltage level. 
Figure \ref{fig:adc_lin} shows the ADC response to DC test input levels. 
The overall response of the ADC approximately follows an sigmoid function, due to saturation effects at both extremes.
The Differential non-linearity (DNL) and Integral non-linearity (INL) are shown in Fig. \ref{fig:adc_dnlinl}. 
Some local non-linearities can also be seen: 
DNL is as large as 10 LSB for some ADC codes, 
and INL is much worse at the margin of ADC codes than the middle region.

\begin{figure}[!htb]
\begin{center}
\includegraphics[width=0.9\textwidth]{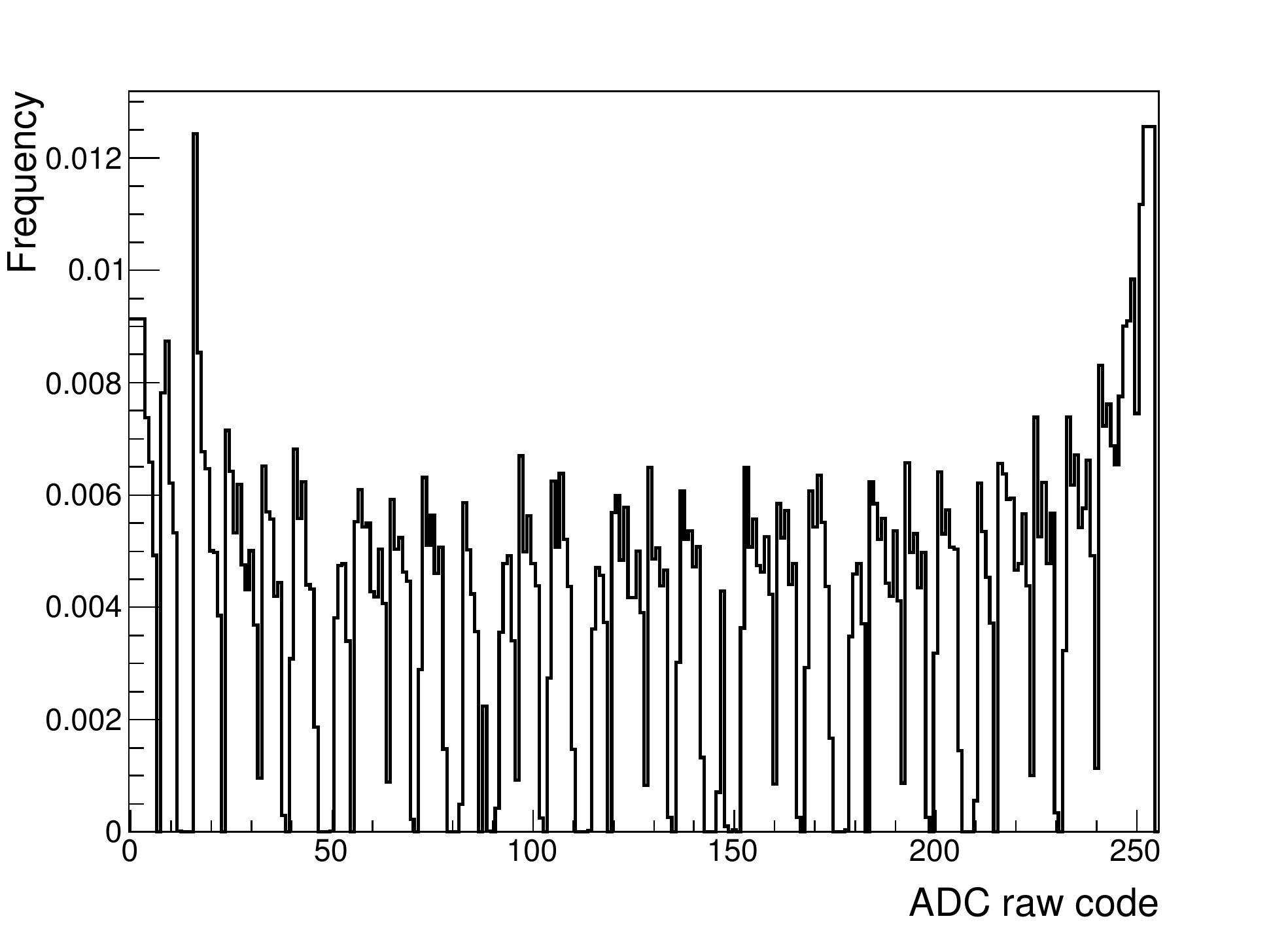}
\caption{ADC code frequencies from a (uniform) triangular wave input.  Missing codes can be clearly seen. The catenary shape of the overall distribution is consistent with the shape in Fig. \protect\ref{fig:adc_lin}.}
\label{fig:adc_codefreq}
\end{center}
\end{figure}

To study these non-linearity in detail, we provided a slow ramp, 2\,Hz triangular voltage wave to the test input. 
The triangle wave amplitude and offset were set to exclude the peaks from the digitization range, 
making the input voltage distribution uniform across the full ADC range.  Because the input voltage was uniform, we expect
the output code distribution to also be uniform, if the ADC were linear.
The measured frequency distribution of codes is shown in Fig. \ref{fig:adc_codefreq}. 
The code frequency distribution is non-uniform, and some codes have zero frequency, which demonstrates
non-linearity above 1 LSB.

Detailed circuit simulations indicate that the ADC code distortions are most likely caused 
by a larger-than-expected mismatch in the capacitor values used in DASC and ADSC.
This mismatch can cause gain errors in the ADC stages which limit the resolution (before calibration).
To reduce power consumption, POM used very small capacitors, outside the range used by the foundry
when quoting variability.  Previous experience with other processes indicated that the
quoted variability would scale beyond the quoted range, but this did not prove to be the case.
Larger capacitors could be used to reduce this non-linearity, at the cost of additional power consumption.
Also from the simulation study, the lower-than-normal operation clock for ADC performance evaluation 
was chosen where the best performance was expected despite of the capacitor mismatch.

The ADC linearity can be improved by offline calibration.  We
calibrated the  ADC response using the measured code frequency distribution from the uniform
input. 
To avoid edge distortions, the frequency of codes less than 3 or greater than 252
are taken to be those of their nearest measured neighbor.
We define a corrected code using the following formula:
\begin{equation}
c(n) = 256 \times \sum_{i=0}^{n-1} f(i)~~\rm{,}
\label{eq:corrected_code}
\end{equation}
where $n$ is the raw code, $f(n)$ is the measured code frequency, and $c(n)$ is the corrected code ($c(0) \equiv 0$).
The arbitrary scale factor of 256 allows convenient comparison with the uncorrected codes.

\begin{figure}[!htb]
\begin{center}
\includegraphics[width=\textwidth]{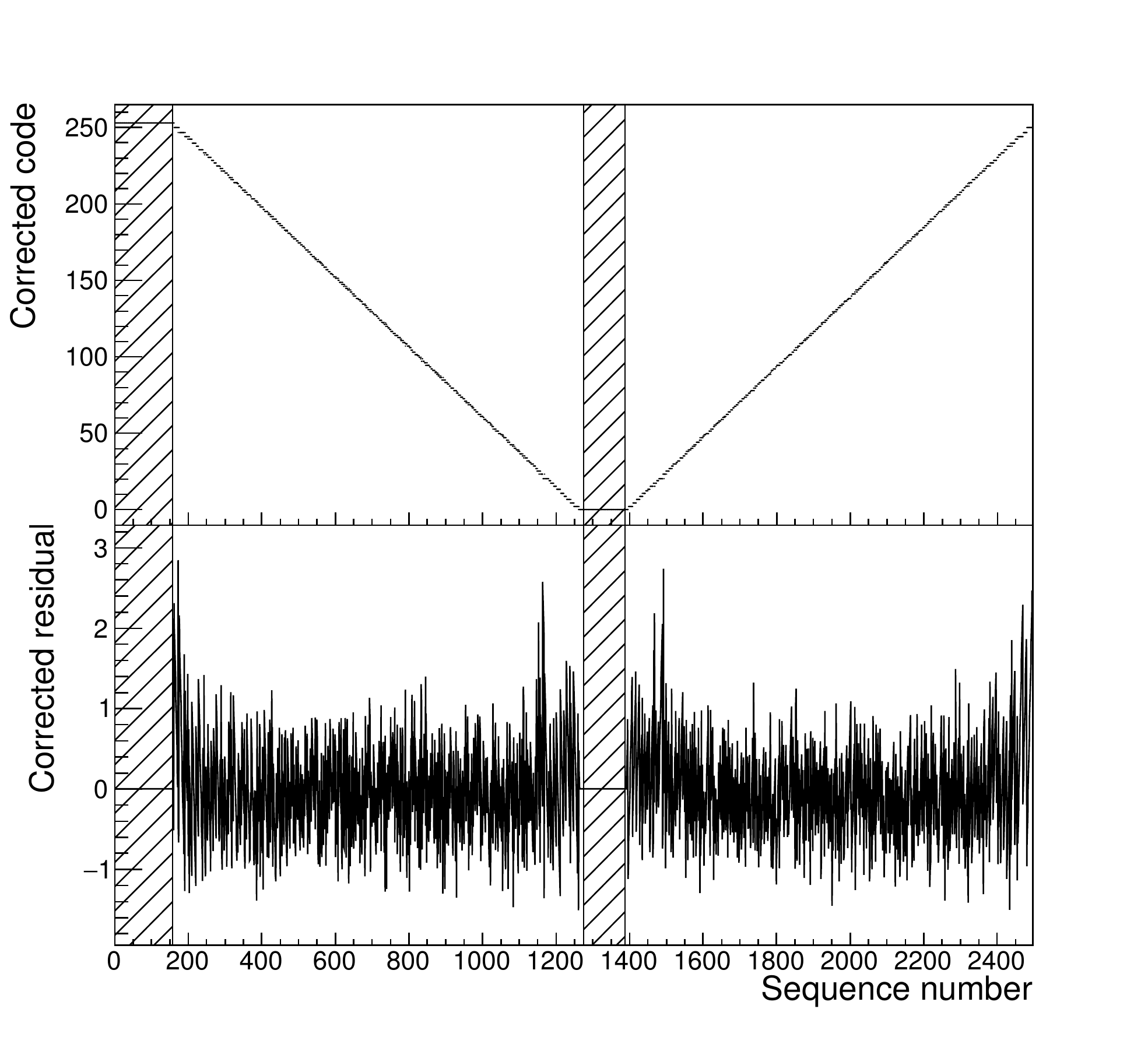}
\caption{Top: Corrected code (black dots.)
Bottom: Fit residuals. We do not include the hatched region in the fit since it is biased by input overflows.
}
\label{fig:adc_fit}
\end{center}
\end{figure}

To evaluate the quality of the code correction, 
we plot the corrected code response given an independent sample 
of the same triangular wave input, and fit it to a truncated triangular wave. 
The response, the fit, and the fit residuals are shown in Fig. \ref{fig:adc_fit}. 
The truncated triangular function has four fit parameters: 
period, amplitude, and mean code value of the non-truncated triangular wave, 
as well as the maximum code of the ADC. The error of each corrected code is set to 0.6 LSB
in the fit,
which yields a $\chi^2$ per degree of freedom (NDF) of 2214/2235, consistent with unity.

The root-mean-square (RMS) of the fit residuals after code
corrections is measured to be 0.611 LSB.  From this,
we calculate the Effective Number of Bits (ENOB) performance of the calibrated POM ADC using~\cite{ref:ieee1057}.
  \begin{equation}
    \rm{ENOB} = 8-\log_2(\sqrt{12}\times \rm{RMS}) = 6.9~.
    \label{ENOB}
  \end{equation}
A similar calculation using the uncorrected codes yields an ENOB of 4.6.

The residual RMS after the code corrections combines remaining
nonlinearities and noise fluctuations.  
The noise on a single sample can be estimated as $\sqrt{0.5}$ times the
RMS of the difference between two independent samples measured with a
constant voltage input, which we determined to be 0.56 LSB. 
By subtracting this noise contribution in quadrature from the fit residual RMS, 
we estimate the effect of uncorrected non-linearities to be 0.24 LSB. 
Thus the code frequency calibration recovers most of the intrinsic precision of the POM ADC.

\begin{figure}[!htb]
\begin{center}
\includegraphics[width=0.9\textwidth]{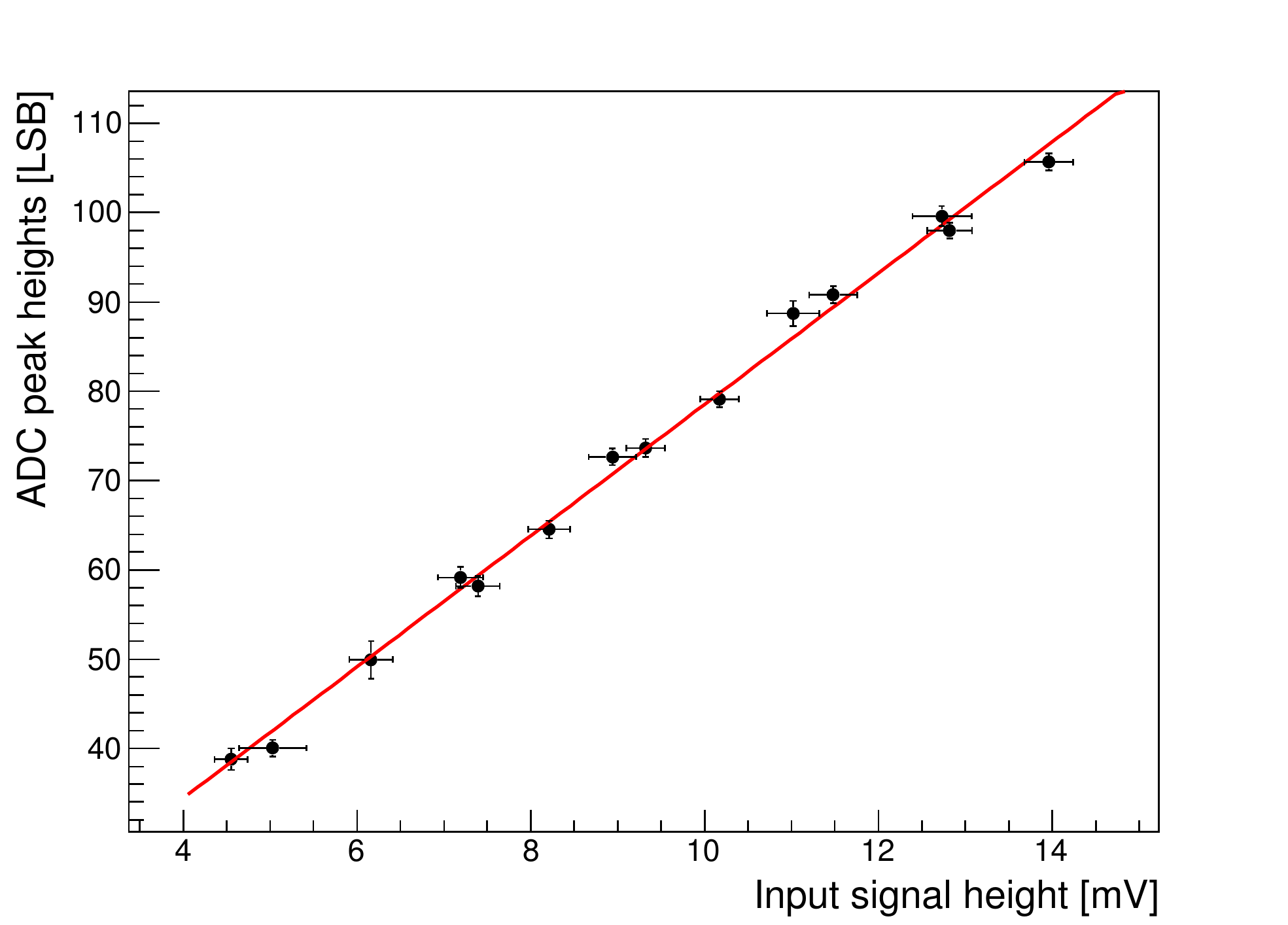}
\caption{Black dots: The linearity relation between baseline-subtracted ADC peak (vertical axis)  and input signal height (horizontal axis). Solid line: straight line fit with  $\chi^2/\rm{NDF} = 0.403$~.}
\label{fig:adc_anal_lin}
\end{center}
\end{figure}

The ADC waveform performance was studied using the analog voltage
inputs (routed to the receiver buffers) similar to the actual detector signals,
with a rise/fall time of $6\,\rm{ns} / \sim\,100\,\rm{ns}$ and an amplitude of $3 \sim 15\,\rm{mV}$.
The two pre-trigger ADC samples were used to measure the baseline (average) and noise level (difference). 
The baseline was subtracted from the eight post-trigger ADC samples, then the signal charge is reconstructed from the integral or the peak of baseline-subtracted ADC samples.  
The linearity relation between baseline-subtracted ADC peak and input signal height is shown Fig.~\ref{fig:adc_anal_lin}. 
A straight line fit with $\chi^2/\rm{NDF} = 0.403$ was found.  
From the pre-trigger samples, the per-sample noise level was measured  to be
2\,LSB, or 1.6\,\% of the total dynamic range, or 5.5\,mV at the ADC input. 

\subsection{TDC Performance}

\begin{figure}[!htb]
\begin{center}
\includegraphics[width=0.9\textwidth]{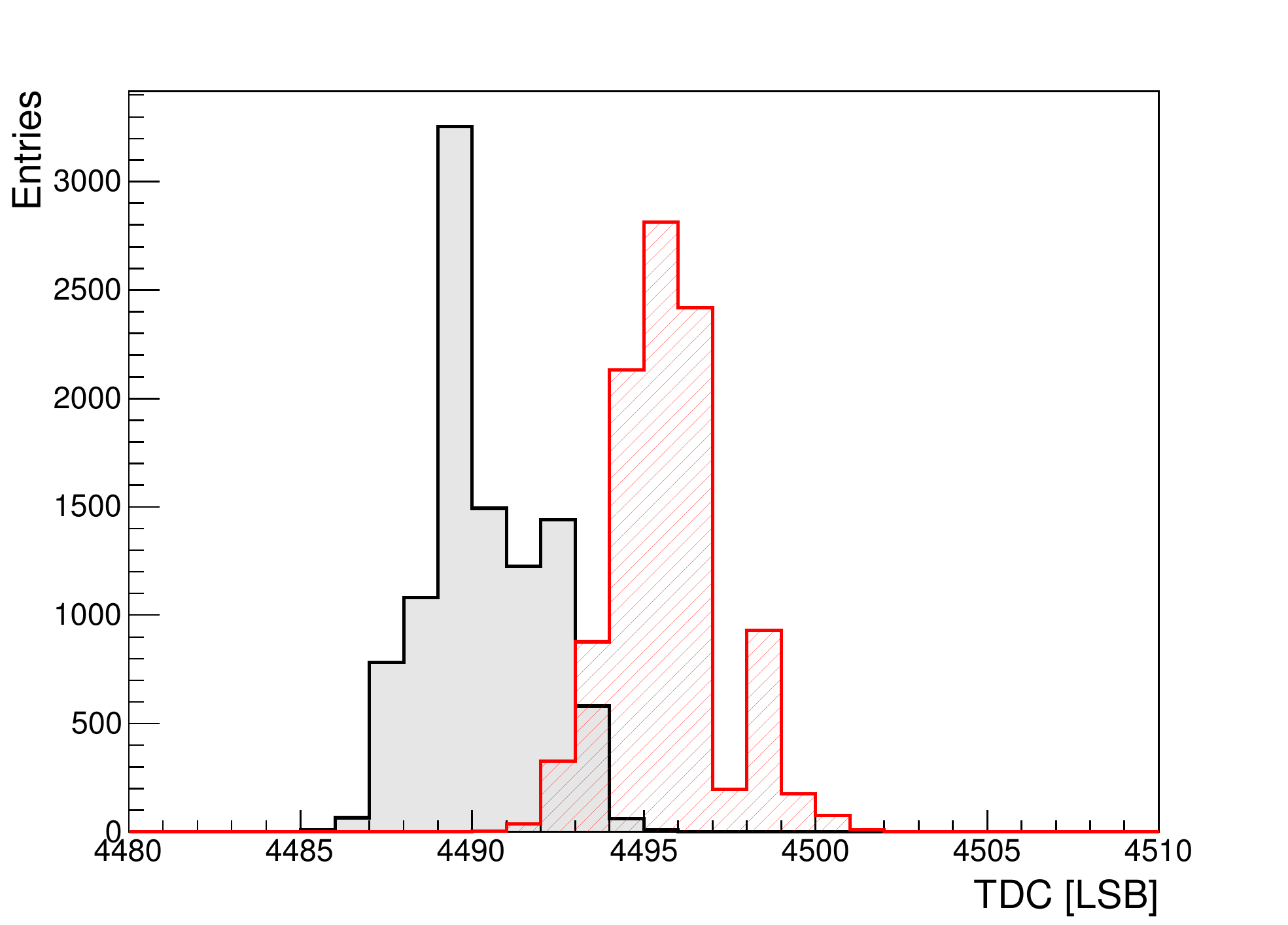}
\caption{Raw TDC distributions with a fixed delay between the TDC
  start and stop signals for a single POM channel. The RMS of TDC0
  (Gray-filled histogram) and TDC1 (slash-line-filled histogram) are
  1.7 and 1.6 LSB, respectively. The mean values of both histograms
  come from an arbitrary signal delay setting.}
\label{fig:tdc_tdc}
\end{center}
\end{figure}

In order to measure the absolute time resolution of the TDC, we used a pulse generator to send both the start and stop signals
to the TDC test inputs, using a passive splitter and several meters of  
cable to delay the stop. 
The distribution of the measured TDC values is shown in Fig. \ref{fig:tdc_tdc}.
The RMS width of the distribution is found to be $1.7$ TDC counts, which corresponds to 67\,ps after calibration (described below).
This resolution includes the input signal jitter, which we estimate to be $< 25\,\rm{ps}$.
Note that this is the optimum resolution, only valid for very short delays between start and stop.  Resolution as a function of delay is discussed below.

\begin{figure}[!htb]
\begin{center}
\includegraphics[width=0.9\textwidth]{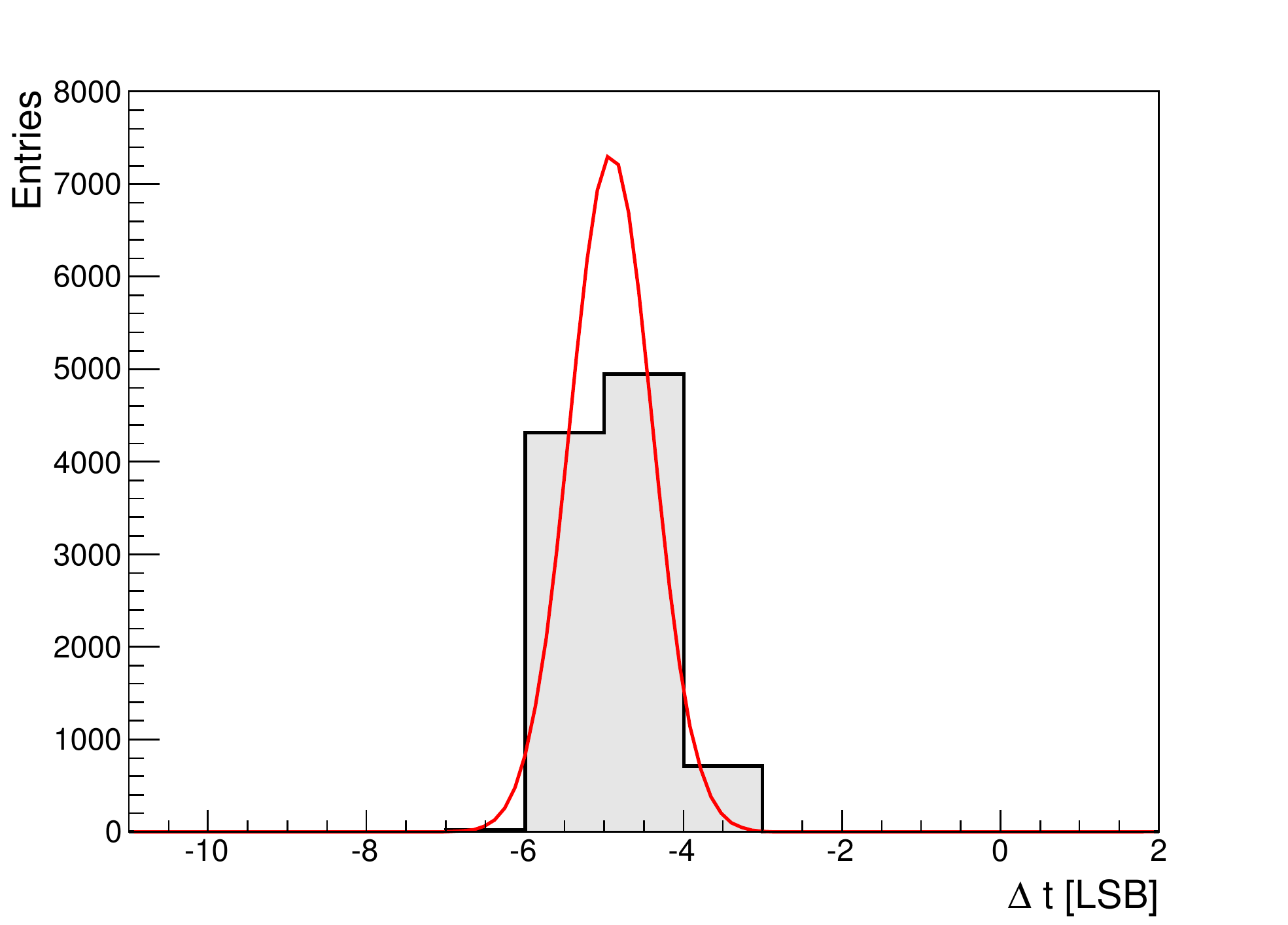}
\caption{The $\Delta t = \rm{TDC0} - \rm{TDC1}$ distribution (filled
  histogram) with a fixed delay for a single
POM channel, and its fit to the Gaussian distribution (solid line) with fit width of 0.5 LSB. 
The -5 LSB offset of $\Delta t$ is purely artificial effect coming from the difference of path length to TDC0 and TDC1.
}
\label{fig:tdc_dt}
\end{center}
\end{figure}

The difference between the two TDC measurements in a single channel ({\deltat} $\equiv \rm{TDC0} - \rm{TDC1}$)
for the fixed delay test configuration is shown in 
Fig. \ref{fig:tdc_dt}.  The intrinsic {\deltat} resolution is
0.5 TDC counts, which corresponds to 19\,ps after calibration.
This result is consistent with simulation studies, which predicted
0.4 LSB {\deltat} RMS jitter due to quantization effects in small {\deltat} measurement. 
The POM {\deltat} resolution is expected to be better than the absolute time resolution, as the TDCs share
the same oscillator clock, allowing cancellation of any oscillator start-up phase noise.

\begin{figure}[!htb]
\begin{center}
\includegraphics[width=0.9\textwidth]{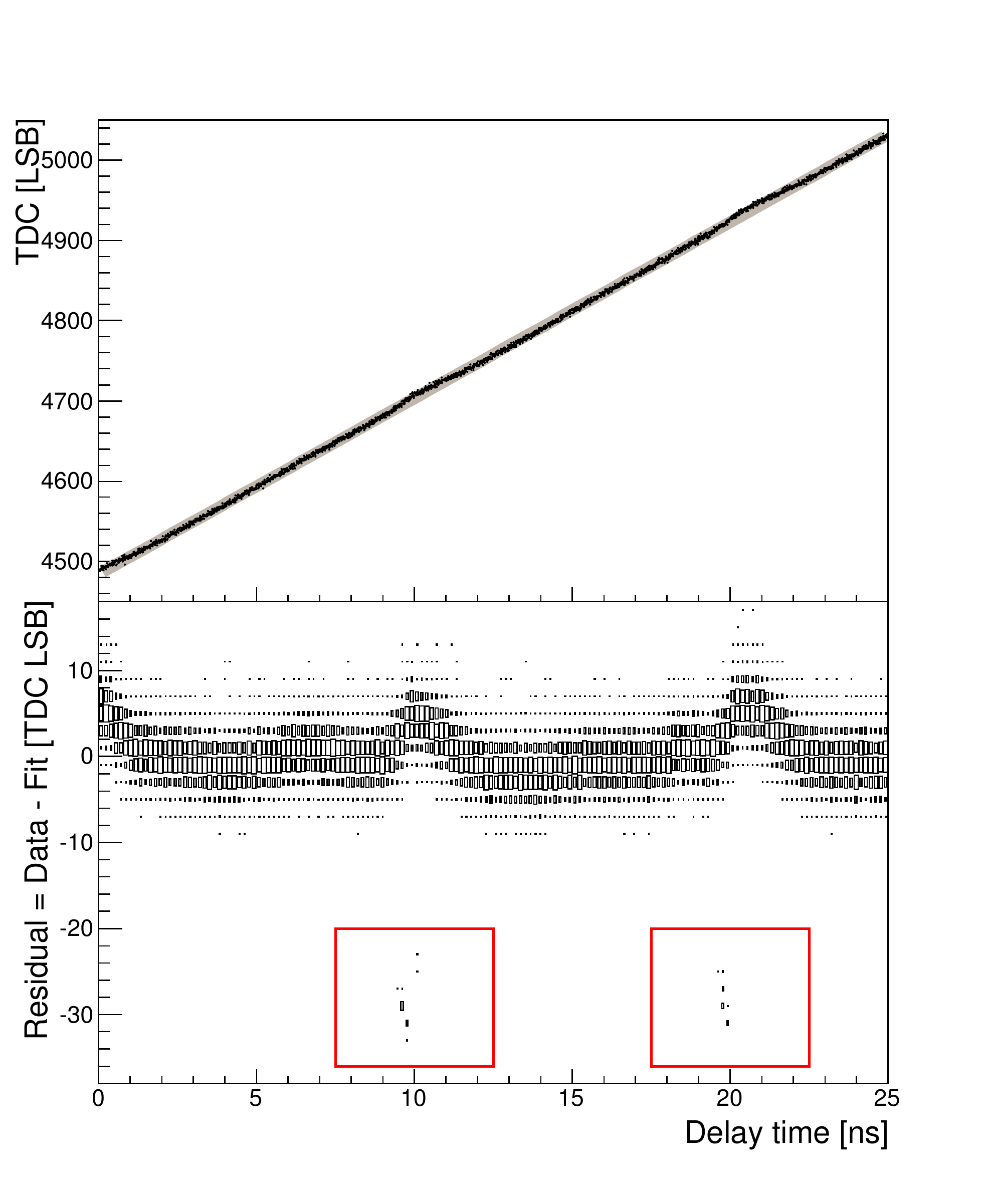}
\caption{Top: Fine scan of the TDC values versus absolute time for
  TDC0 over one clock cycle. 
  The black dots and thick gray line are data and linear fit, respectively. 
  Only 1/1000 data points are drawn for clarity.
  Bottom: Residuals between the data and
  the linear fit. Outlier measurements are emphasized by the solid line boxes.} 
\label{fig:tdc_lin}
\end{center}
\end{figure}

The TDC linearity was measured by scanning the test input stop pulse using a Phase Locked Loop (PLL) generated by
the FPGA.  A fine scan with a 78\,ps step allowed studying the output with up to 25\,ns delay with respect to the
TDC start. 
The data are shown in the upper plot of Fig.~\ref{fig:tdc_lin}, with an overlaid linear fit.  

The bottom plot of Fig.~\ref{fig:tdc_lin} shows the residuals (data -  linear fit) of this fit as a function of the delay.
A small fraction (less than 0.3\%) of the measurements are $\sim$32 counts off from the core residual distribution.
These outliers appear in distinct clumps separated by 10\,ns, with slightly different timing for TDC0 and TDC1,
roughly synchronous with the leading and trailing edges of the 50\,MHz
operation clock used by the ADC and the digital backend.
When we configure the FPGA to disable the operation clock during the TDC digitization interval, the outliers disappear.
We interpret these results as due to the digitally induced noise in the TDC. 
From simmulation study, it is found that
the most likely crosstalk path is through the power and ground of the digital supplies. 
A reduction in the effective supply voltage due to the power surges occurring at the clock transitions
could induce delays that cause
the incorrect capture edge to be selected for the ring cycle counter. 
The TDC digital noise could be reduced by 
optimizing both the pad ring and the supply distribution network. 

\begin{figure}[!htb]
\begin{center}
\includegraphics[width=0.9\textwidth]{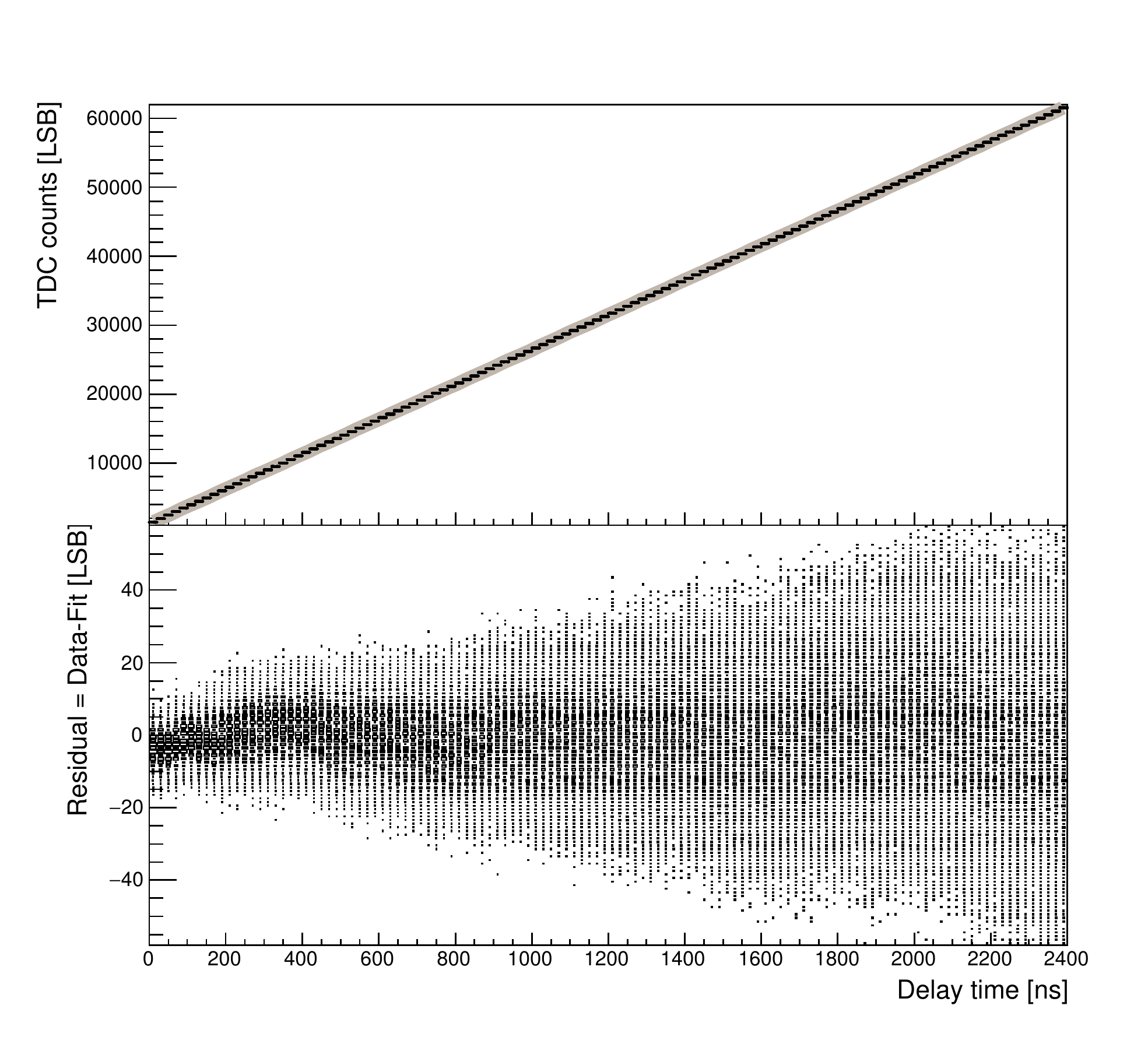}
\caption{Top: Coarse scan of TDC versus absolute time delay over the full dynamic range. Black dots and gray thick line are data and linear fit, respectively. Bottom: Residual between data and linear fit. }
\label{fig:tdc_lin_all}
\end{center}
\end{figure}

The TDC response over the full $2 {\mu}\rm{s}$ time window was measured by a coarse FPGA PLL
scan of the test input stop delay with a 20\,ns delay step. 
The data and a superimposed linear fit are  shown in the upper plot of Fig.~\ref{fig:tdc_lin_all}.  
The data show good linearity, and
we use the slope of this
line to measure the TDC LSB to be 37\,ps per TDC count. 
Outlier measurements do not show up in this test due to the coarse scan step.
The bottom plot of Fig. \ref{fig:tdc_lin_all} shows the residual (TDC
data - linear fit), in which an increasing spread as a function of the
delay time is apparent.  
The absolute TDC resolution, 
which is defined as the RMS width of TDC distribution for a fixed time delay set,
increases from 2 LSB (at 0 time delay) 
to 13 LSB ($2 {\mu}\rm{s}$ time delay) over this range.
The worst case resolution of 13 LSB corresponds to 480\,ps, 
which still meets the POM requirements.
 
Detailed circuit simulations show that the observed increase of the
resolution  with the delay is
due to accumulating random phase noise in the ring oscillator and the
PLL of the FPGA.
Phase noise could be reduced
by phase-locking the ring oscillator to the external reference clock. 

\subsection{Discriminator Performance}

\begin{figure}[!htb]
\begin{center}
\includegraphics[width=0.9\textwidth]{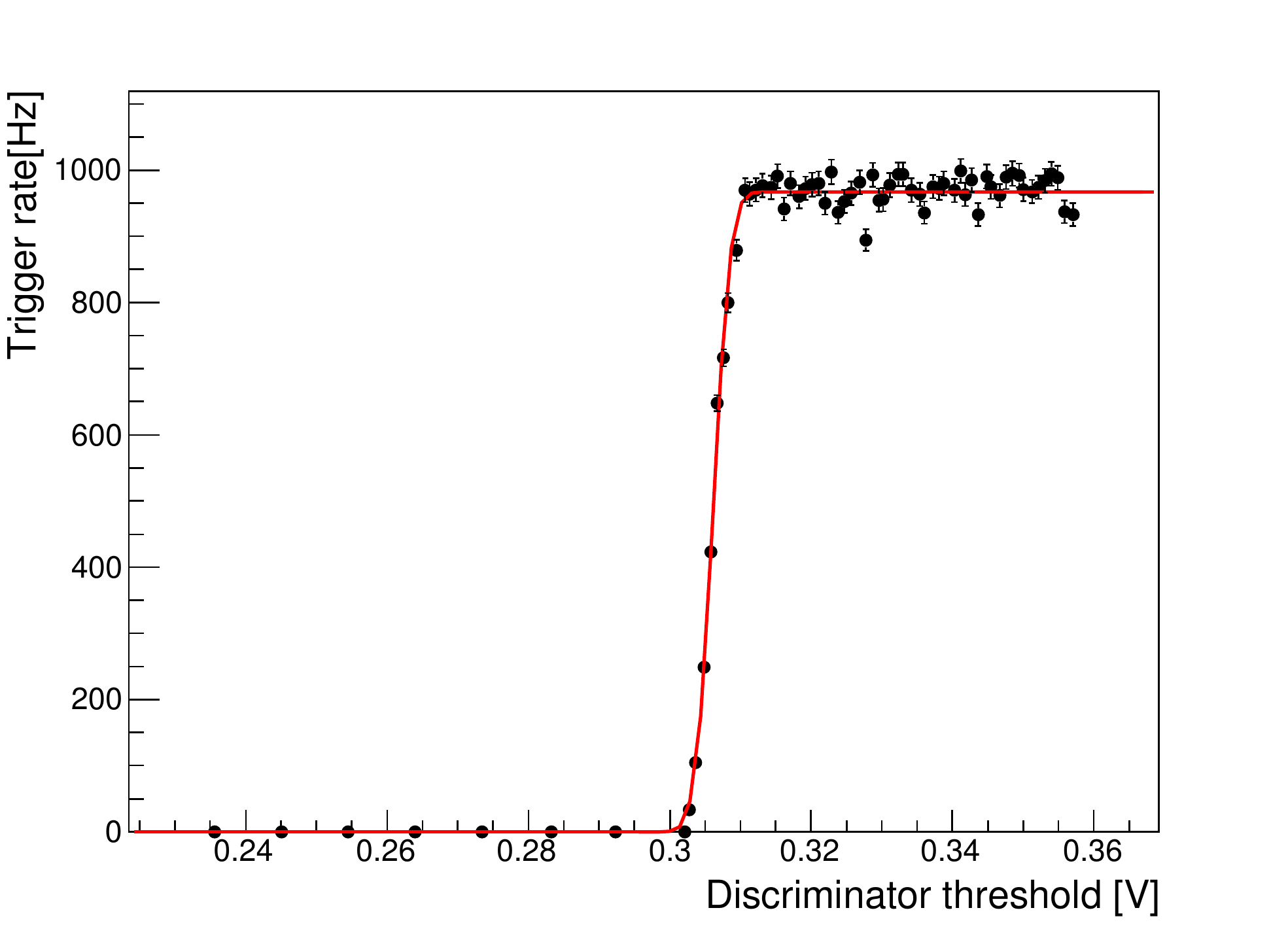}
\caption{The trigger rate of the POM versus the discriminator
  threshold (black dots) and the error function fit (solid red line). } 
\label{fig:disc_vthscan}
\end{center}
\end{figure}

The discriminator turn-on curve is shown in Fig.~\ref{fig:disc_vthscan}.  
This was measured by scanning the discriminator threshold for a fixed input pulse, and measuring the trigger rate.  
The maximum trigger rate shown at high thresholds is
consistent with the input pulse rate. 
The discriminator thus transits from off to full efficiency in 10\,mV,
with normalized discriminator gain of 0.12\,/mV,
and discriminator resolution of 2.7\,mV, which are estimated from fit.
Simulation studies predict a discriminator jitter of around 0.5\,mV, 
which indicates a substantial additional noise on the input or threshold voltage.
A scan of the input pulse height yielded consistent turn-on curves, 
and showed a roughly linear dependence of the threshold value on the input pulse height.

\begin{figure}[!htb]
\begin{center}
\includegraphics[width=0.9\textwidth]{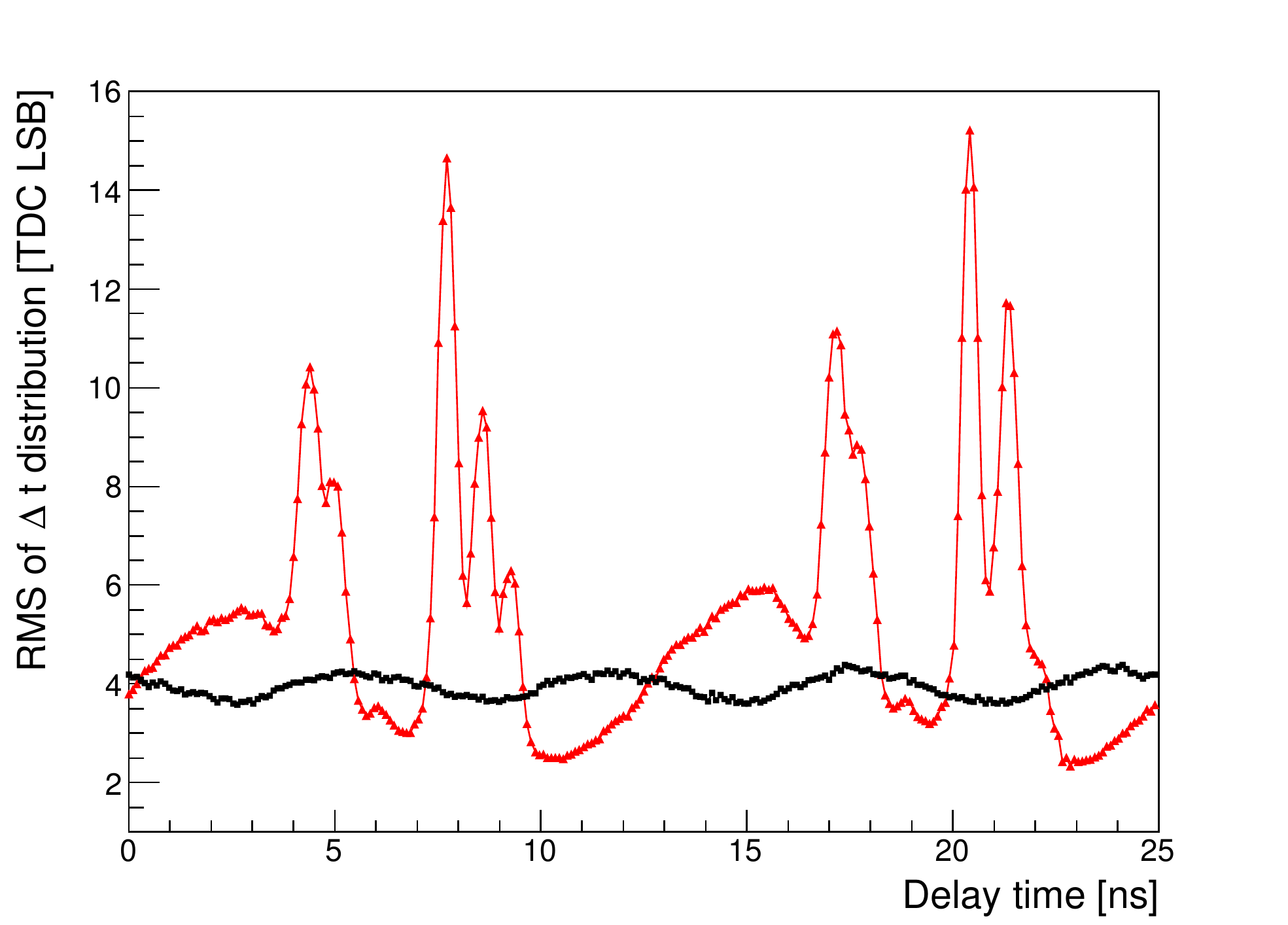}
\caption{The RMS of the TDC $\Delta t$ as a function of delay within
  one clock phase. The red triangular dots and square black dots
  correspond to the system clock on and off transitions, respectively. }
\label{fig:disc_dt}
\end{center}
\end{figure}

Due to the finite ($\sim 10\,{\rm ns}$) rise time of the input signals, the discriminator introduces an amplitude dependent offset (latency) to the time measurement,
{\it i.e.} time walk effect.
Fluctuations in the discriminator threshold or the signal baseline degrade the time resolution.  The effect of the 
discriminator on the POM time resolution is shown in Fig.~\ref{fig:disc_dt} (red triangle graph).  For this test, a voltage pulse emulating
the expected wire signal, delayed by the FPGA PLL, was split and routed through the receiver buffers
and then to the two POM discriminators.  The discriminator
outputs were then used to stop the TDCs. 
The plot shows the root-mean-square (RMS) spread of the TDC {\deltat},
as a function of the delay value.  Under normal operating conditions, a large resolution spread is seen,
with the magnitude of the spread depending on the precise delay of the signals, synchronous with the system clock.  Under these
conditions, the overall RMS of the {\deltat} distribution was measured to be 10 TDC LSB, or 370\,ps, which
does not meet the POM specifications.

Also shown in Fig.~\ref{fig:disc_dt} (black square graph) is the TDC difference for this same setup, but where the system clock is disabled
during the TDC sampling period.  In this case, the RMS is essentially constant at 4 TDC LSB, or 150\,ps.
These results show that
the discriminator input-referred differential noise exceeds specifications, with the dominant noise originating from the feed-through of the system clock to the threshold voltage. 
A simulation of the noise coupling to the discriminator reference
voltage through the mutual inductance of the bond wires demonstrated
effects consistent with those measured in the POM prototype.  
Coupling between the discriminator and the system clock could be reduced
by buffering all reference voltages internal to the IC.

\subsection{Power consumption}

POM has separate supply voltages for the ADCs, TDCs, preamplifiers, and the digital backend. 
The power consumption of each circuit was determined by measuring the currents drawn by
those supplies using the normal operating configuration with a 1\,kHz
trigger rate. 
Unused circuits can be selectively turned off to reduce the power consumption.
Table \ref{table:power} summarizes the power consumption of 4 channels of POM. 
The analog parts of the TDCs and ADCs consume only 1.3\,mW and 2\,mW per channel, respectively. 
The total power consumption per POM channel, including the digital backend, is 32\,mW
for the internal signal amplification, and 26\,mW for the external signal
amplification, meeting the requirements.
\begin{table}[!htb]
\caption{POM power consumption for 4 channel operation. 
Note that the total power consumption is not a simple sum of all entries: for internal / external preamplifier mode, receiver buffer / preamplifier is not used, respectively.
}
\begin{center}
\begin{tabular}{l c}
\hline
Part & Power consumption (mW)  \\
\hline
Preamplifiers & 50 \\ 
ADC (Analog part only) & 8 \\ 
TDC (Analog part only) & 5 \\ 
LVDS driver & 22 \\
Receiver buffer & 28 \\ 
Digital backend & 41 \\ 
\hline
Total (internal / external preamplifier) & 126 / 104 \\
Total per channel (internal / external preamplifier) & 32 / 26 \\
\hline
\end{tabular}
\end{center}
\label{table:power}
\end{table}

\subsection{Radiation tolerance}

Neutrons produced by the muon capture on the nucleus are the dominant 
radiation source expected for the Mu2e tracker electronics.
Neutron scattering can deposit energy within the active volume of the CMOS components, causing non-destructive
state changes in the transistors (Single Event Upsets: SEUs).  Neutron
capture followed by beta decay can transmute silicon to phosphorus,
changing the doping and permanently damaging the circuit (Type Inversion). 
Based on detailed \textsc{Geant4} simulations \cite{ref:Geant4}, the
total neutron dose at the location of the frontend electronics in Mu2e is
expected to be $2 \times 10^{12}\,\rm{n}/\rm{cm}^2$, integrated over 3 years.

The ADC occupies the largest area of the POM.  The ADC is also expected to have the largest sensitivity to
the long-term neutron exposure (total dose) compared to the other POM components.
We have tested the tolerance of the HIPPO ADC \cite{ref:HIPPO}, on
which the POM ADC design is based, to neutron irradiation at the 
MNRC reactor facility in McClellan, CA.  The reactor neutrons have a similar energy spectrum as those
 expected for Mu2e.
 An unbiased HIPPO chip was exposed to the
 total integrated neutron dose of $2 \times 10^{14}\,\rm{n}/\rm{cm}^2$, which is 
100 times larger than the expected total neutron dose for an ADC in Mu2e.
The performance of the HIPPO ADC after irradiation in terms of noise and linearity was
found to be consistent with the pre-radiation performance, showing no degradation.

Triple-redundant circuitry was used in the configuration register of
the POM to minimize the SEU sensitivity, but
not in the core ADC and TDC or the data transmission buffers.
An attempt to directly measure the SEU rate of the POM in a neutron flux of
approximately $1.5 \times 10^{10}\,\rm{n}/\rm{cm}^2/s$ was inconclusive, as the 
ethernet signal driver chip on the test board failed after only a few seconds of radiation exposure. 

\section{Summary and Conclusions}

We developed and tested a prototype ASIC digitizer for integrated wire chamber
readout in a single chip, implemented in 65\,nm commercial CMOS technology.
The specifications targeted the requirements for the readout of the Mu2e straw tracker.
The prototype ASIC TDC was found to meet or exceed 
the design goals, in spite of
a small rate of TDC outliers.
The ADC was found to meet the design goals only when operated in slower frequency,
with some non-linearity effects.
Simple design changes should improve the TDC ring oscillator latching and the ADC
linearity.
The discriminator, and therefore the analog signal operation mode of the ASIC,
did not meet the design goals, due to
clock-induced noise on the discriminator threshold.
Improved isolation should reduce this problem.

The Mu2e digitizer electronics requirements have changed
since the POM prototype ASIC was designed; in particular, 
increased cooling capacity has relaxed the power consumption
requirement by an order of magnitude.  The new requirements
can be met by existing commercially available components (ADC
and FPGA).  Mu2e has consequently decided to implement the
digitizer using commercial components, and no further development
of the POM ASIC is foreseen.
 
POM is one of the first ASICs developed for High Energy Physics
in a 65\,nm CMOS process.  This process will be valuable for future
HEP detector readout applications requiring
compact circuits, low-power, reliability, and radiation hardness.
The POM prototype showed that
extremely low power consumption can only be achieved when
precision matching of components is not required, or the effects of
component mismatch can be corrected by calibration.
It also showed that avoiding cross-talk in such extremely dense circuits requires
careful isolation of analog signals, digital signals, power, ground, and
reference inputs, both in the ASIC and its packaging.

\section*{Acknowledgments}
This material is based upon work supported by the US Department of
Energy (DOE), Office of Science under contract numbers DE-AC02-05CH11231
and DE-AC02-07CH11359.
We thank the personnel of   
the MNRC reactor facility operated by the University of California at
Davis for their support during the radiation tolerance testing.

\end{document}